\documentclass[conference]{IEEEtran}
\usepackage{amsmath,amssymb,bm,cite,algorithm,algpseudocode,amsthm}
\usepackage{graphicx, xcolor}
\usepackage[absolute]{textpos}

\begin{document}
\title{Position Aided Beam Alignment for Millimeter Wave Backhaul Systems with Large Phased Arrays}
\author{
\IEEEauthorblockN{George~C.~Alexandropoulos}
\IEEEauthorblockA{Mathematical and Algorithmic Sciences Lab, Paris Research Center, Huawei Technologies France SASU, \\20 Quai du Point du Jour, 92100 Boulogne-Billancourt, France}
email: george.alexandropoulos@huawei.com}

\begin{textblock}{14}(1,.1)
	\begin{center}
		Extended version of an invited paper at IEEE CAMSAP 2017.
	\end{center}
\end{textblock}

\maketitle
\begin{abstract}
Wireless backhaul communication has been recently realized with large antennas operating in the millimeter wave (mmWave) frequency band and implementing highly directional beamforming. In this paper, we focus on the alignment problem of narrow beams between fixed position network nodes in mmWave backhaul systems that are subject to small displacements due to wind flow or ground vibration. We consider nodes equipped with antenna arrays that are capable of performing only analog processing and communicate through wireless channels including a line-of-sight component. Aiming at minimizing the time needed to achieve beam alignment, we present an efficient method that capitalizes on the exchange of position information between the nodes to design their beamforming and combining vectors. Some numerical results on the outage probability with the proposed beam alignment method offer useful preliminary insights on the impact of some system and operation parameters.
\end{abstract}


\section{Introduction}\label{sec:Intro}
Millimeter wave (mmWave) communication \cite{Rappaport_Access} is a promising technology for addressing the high throughput requirement for the fifth generation (5G) mobile networks \cite{Boccardi_COMmag, Andrews_JSAC}. Short-range mmWave communication at the unlicensed band of 60 GHz is already standardized in IEEE 802.11ad \cite{J:IEEE_11ad} and initial theoretical investigations on mmWave cellular systems \cite{Rappaport_Access, Alkhateeb_JSTSP} have identified their potentials and key challenges. The mmWave frequencies have been also recently considered for the wireless backhaul communication of small cells \cite{Chia_backhaul, Ge_backhaul}, which are expected to be densely deployed as an efficient means for increasing the geographic spectrum reusability \cite{Bhushan_densification}.

Reliable mmWave backhauling depends on very directional communication, which are implemented in practice either with antennas having large apertures or with large phased antenna arrays \cite{Chia_backhaul}. By exploiting the fact that the wavelength at very high frequencies is very small, large phased antennas can be cheaply packed into small form factors and, thus, have been effectively used in realizing highly directional BeamForming (BF) supporting long outdoor links. To achieve the full benefit from BF in a communication link between multi-antenna nodes, the entire channel state information needs to be available at both communication ends. However, this information is hard to acquire in mmWave systems due to the low coherence time, the Radio Frequency (RF) hardware limitations, and the small Signal-to-Noise Ratio (SNR) before BF. Although recent theoretical works \cite{Alkhateeb_JSTSP, Marzi_TSP2016, Bazzi_SPAWC2016, Alexandropoulos_GC16} capitalized on the spatial sparsity of mmWave channels \cite{Zhang_WCNC2010} to estimate portions or the entire channel gain matrix, the presented approaches required lengthy training phases to estimate the channel in multiple directions using complex compressed sensing algorithms. Another family of approaches (e$.$g$.$, \cite{Hosoya_TAP2015, Wang_JSAC, Jeong_COMmag2015, Barati_2015, Raghavan_JSTSP, Hur_TCOM}) for efficient BF is based on beam switching between the communicating nodes in order to find pair(s) of beams from their available codebooks meeting a predefined performance threshold. When such beam pair(s) is(are) found, beam alignment is considered to be achieved and no further beam searching is needed.   

In this paper, we focus on the mmWave backhaul system of \cite{Hur_TCOM} and present a robust beam alignment method for wireless channels including a Line-Of-Sight (LOS) component. The proposed method intends at achieving beam alignment in at most two rounds of control information exchange. The core idea of the method lies on the control information, which is considered to be the new positions of the nodes after each declared beam misalignment event, on its exchange, and on its utilization in designing the BF technique at all communication ends. The position information of a node is assumed to be available to it via an attached position sensor. It is noted that positioning sensors have been also considered in \cite{Maiberger_IEEEI2010, Doff_Internal2015}, however, they were adopted for identifying and circumventing beam misalignment only at the transmit node.     

\textit{Notation:} Vectors and matrices are denoted by boldface lowercase letters and boldface capital letters, respectively. The transpose and Hermitian transpose of a matrix $\mathbf{A}$ are denoted by $\mathbf{A}^{\rm T}$ and $\mathbf{A}^{\rm H}$, respectively, ${\rm diag}\{\mathbf{a}\}$ denotes a square diagonal matrix with $\mathbf{a}$'s elements in its main diagonal, whereas $\mathbf{I}_{n}$ ($n\geq2$) is the $n\times n$ identity matrix. The $i$th element of $\mathbf{a}$ and the $(i,j)$th element of $\mathbf{A}$ are denoted by $[\mathbf{a}]_i$ and $[\mathbf{A}]_{i,j}$, respectively, and $||\mathbf{A}||_{\rm F}$ gives the Frobenius norm of $\mathbf{A}$. $\mathcal{C}$ represents the complex number set, ${\rm card}(\mathcal{F})$ is the cardinality of set $\mathcal{F}$, $|\cdot|$ denotes the amplitude of a complex number, and $\mathbb{E}\{\cdot\}$ is the expectation operator. ${\rm d}(M,N)$ denotes the length of the line segment connecting the points $M$ and $N$. Notation $x\sim\mathcal{C}\mathcal{N}\left(0,\sigma^{2}\right)$ indicates that $x$ is a circularly-symmetric complex Gaussian random variable with zero mean and variance $\sigma^{2}$, while $x\sim\mathcal{U}\left(\alpha,\beta\right)$ represents a uniformly distributed random variable in $[\alpha,\beta]$. 

\section{System and Channel Models}\label{sec:System_Model}
\begin{figure}[!t]
\centering
\includegraphics[width=0.37\textwidth]{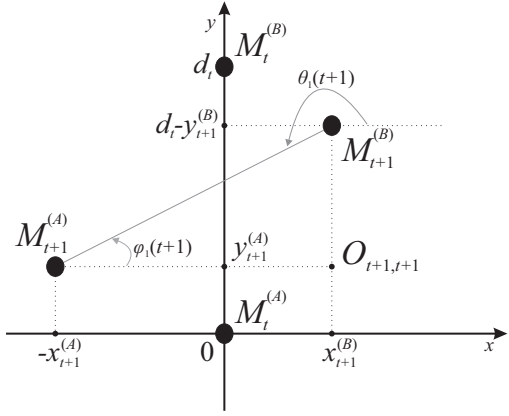}
\caption{The considered point-to-point wireless backhaul communication system. The positions of the nodes $A$ and $B$ at the time instant $t$ are denoted by the points $M_t^{(A)}$ and $M_t^{(B)}$, respectively, and define a Cartesian coordinate system. At the time instant $t+1$ both nodes $A$ and $B$ move to new positions represented by the points $M_{t+1}^{(A)}$ and $M_{t+1}^{(B)}$, respectively. These new positions determine the coordinates of the point $O_{t+1,t+1}$; its first subscript refers to $M_{t+1}^{(A)}$ and the second to $M_{t+1}^{(B)}$.}
\label{Fig:BA_figure}
\end{figure}

Suppose a wireless backhaul communication system operating in the mmWave frequency band and consisting of two half duplex multi-antenna transceiver nodes $A$ and $B$ (see Fig$.$~\ref{Fig:BA_figure}). The fixed positions of the nodes defines a Cartesian coordinate system, according to which the positions of nodes $A$ and $B$ at the time instant $t$ are represented by the points $M_t^{(A)}$ and $M_t^{(B)}$, respectively. Their respective coordinates are $(0,0)$ and $(0,d_t)$, where $d_t$ denotes the physical distance between the nodes at this time instant. Regarding the mmWave access, each node is assumed to be equipped with one RF chain and is capable of realizing only analog transmit BF when being in transmit mode, and only analog receive combining when being in receive mode. Node $A$ is assumed to have a large uniform linear antenna array (ULA) with $N_A$ elements, whereas node $B$ is equipped with a large $N_B$-element ULA. We hereinafter assume for simplicity that the positions of the nodes coincide with the positions of the centers of their antenna arrays.

To establish wireless communication between nodes $A$ and $B$, i$.$e$.$, to meet a minimum performance threshold for reliable information exchange, a control phase for channel access is adopted. We assume that communication is realized in discrete time instants each including a control and a data phase. The control phase constitutes of several time slots during which data communication is being set up. We assume that during each discrete time instant the channel remains constant, but it may change between different instances. During the control phase for channel access, control information needs to be exchanged between nodes $A$ and $B$. This control information exchange requires in general a much less stringent performance threshold than data exchange. This means that BF gain is not necessary for control signaling, hence, its received SNR can be afforded to be low. As such, we assume that control signaling exchange can be in principle handled by the considered communication systems. 

Before signal transmission from node $A$, the unit power data stream $s\in\mathbb{C}$ (chosen from a discrete modulation set) is processed by a BF vector $\mathbf{v}_A\in\mathbb{C}^{N_A\times 1}$, and upon signal reception at node $B$, the combining vector $\mathbf{u}_B\in\mathbb{C}^{N_B\times 1}$ is used for processing the received signal. Similarly, node $B$ makes use of the BF vector $\mathbf{v}_B\in\mathbb{C}^{N_B\times 1}$ when transmitting, and node $A$ utilizes the combining vector $\mathbf{u}_A\in\mathbb{C}^{N_A\times 1}$ when being in receive mode. Due to practical limitations with the considered analog antenna arrays \cite{Alkhateeb_JSTSP, Hur_TCOM}, the analog antenna weights at both nodes are supposed to take values from discrete sets. In particular, we assume that $\mathbf{v}_A$ and $\mathbf{u}_A$ at node $A$ are chosen from the finite set $\mathcal{F}_A$ of predefined $N_A$-element vectors. The same holds for node $B$; $\mathbf{v}_B$ and $\mathbf{u}_B$ belong to the finite set $\mathcal{F}_B$ of predefined $N_B$-element vectors. Without loss of generality, it is assumed that for any $\mathbf{f}\in\mathcal{F}_A$ with $\mathbf{f}\in\mathbb{C}^{N_A\times 1}$ and any $\mathbf{z}\in\mathcal{F}_B$ with $\mathbf{z}\in\mathbb{C}^{N_B\times 1}$ holds $|[\mathbf{f}]_{i}|^2\triangleq N_A^{-1}$ with $i=1,2,\ldots,N_A$ and $|[\mathbf{z}]_{j}|^2\triangleq N_B^{-1}$ with $j=1,2,\ldots,N_B$. When node $A$ transmits information to node $B$, the output signal after the combiner at node $B$ can be mathematically expressed as
\begin{equation}\label{Eq:System_Model}
y_B = \sqrt{p}\mathbf{u}_B^{\rm H}\mathbf{H}_{BA}\mathbf{v}_As + \mathbf{u}_B^{\rm H}\mathbf{n}_B,
\end{equation}
where $p$ is the transmit power, $\mathbf{H}_{BA}\in\mathbb{C}^{N_B\times N_A}$ denotes the channel gain matrix between nodes $B$ and $A$, and $\mathbf{n}_B\in\mathbb{C}^{N_B\times 1}$ represents the zero-mean Additive White Gaussian Noise (AWGN) vector with covariance matrix $\sigma_B^2\mathbf{I}_{N_B}$. 

We adopt a geometric channel model with $L$ scatterers similar to \cite{Zhang_WCNC2010, Alkhateeb_JSTSP}. We assume that during each discrete time instant the wireless channel remains constant, but it may change between different instances. According to this channel model, $\mathbf{H}_{BA}$ included in \eqref{Eq:System_Model} is expressed as
\begin{equation}\label{Eq:Channel_Model}
\mathbf{H}_{BA} \triangleq \mathbf{A}_B\left(\boldsymbol{\theta}\right){\rm diag}\{\boldsymbol{a}\}\mathbf{A}_A^{\rm H}\left(\boldsymbol{\phi}\right),
\end{equation}
where $\mathbf{A}_A\left(\boldsymbol{\phi}\right)\in\mathbb{C}^{N_A\times L}$, with $\boldsymbol{\phi}\triangleq[\phi_1\,\phi_2\,\cdots\,\phi_L]$, and $\mathbf{A}_B\left(\boldsymbol{\theta}\right)\in\mathbb{C}^{N_B\times L}$, with $\boldsymbol{\theta}\triangleq[\theta_1\,\theta_2\,\cdots\,\theta_L]$, are defined as 
\begin{subequations}\label{Eq:Angles}
\begin{equation}\label{Eq:A_TX}
\mathbf{A}_A\left(\boldsymbol{\phi}\right) \triangleq \left[\mathbf{a}_A\left(\phi_1\right)\,\mathbf{a}_A\left(\phi_2\right)\,\cdots\,\mathbf{a}_A\left(\phi_L\right)\right],
\end{equation}
\begin{equation}\label{Eq:A_RX}
\mathbf{A}_B\left(\boldsymbol{\theta}\right) \triangleq \left[\mathbf{a}_B\left(\theta_1\right)\,\mathbf{a}_B\left(\theta_2\right)\,\cdots\,\mathbf{a}_B\left(\theta_L\right)\right].
\end{equation}
\end{subequations}
In \eqref{Eq:Angles}, variable $\phi_\ell\in[0,2\pi]$ with $\ell=1,2,\ldots,L$ denotes the $\ell$th path's Angle of Departure (AoD) from node $A$ and variable $\theta_\ell\in[0,2\pi]$ represents the $\ell$th path's Angle of Arrival (AoA) at node $B$. Following the investigations in \cite{Muhi-Eldeen}, we assume that the $1$st channel path is a LOS one with energy much larger than each of the rest $L-1$ paths. In addition, $\mathbf{a}_A\left(\phi_\ell\right)\in\mathbb{C}^{N_A\times 1}$ and $\mathbf{a}_B\left(\theta_\ell\right)\in\mathbb{C}^{N_B\times 1}$ are the array response vectors at nodes $A$ and $B$, respectively (these vectors are given by \cite[eq. (5)]{Alkhateeb_JSTSP} for ULAs). In \eqref{Eq:Channel_Model}, $\boldsymbol{a}\in\mathbb{C}^{L\times 1}$ includes the channel path gains $\alpha_\ell$ $\forall\ell=1,2,\ldots,L$. We further assume that each path's amplitude is Rayleigh distributed and, in particular, that each $\alpha_\ell\sim \mathcal{CN}(0,N_AN_BP_L)$, where $P_L$ denotes the average pathloss between nodes $B$ and $A$.  

In practical deployments of wireless backhaul systems, the antenna arrays of the communicating nodes are usually mounted on outdoor structures that are exposed to wind flow and gusts. These structures are susceptible to movement (or sway) due to wind or ground vibration, which might cause unacceptable Outage Probability (OP) if beam alignment is not frequently performed \cite{Hur_TCOM}. To capture the random movements of the antenna arrays at both nodes $A$ and $B$, we model the displacements at the $x$-axis and $y$-axis for both of them as 
\begin{subequations}\label{Eq:AB_movement}
\begin{equation}\label{Eq:A_movement}
x^{(A)}\sim\mathcal{U}(-x^{(A)}_{\rm w},x^{(A)}_{\rm e}),\, y^{(A)}\sim\mathcal{U}(-y^{(A)}_{\rm s},y^{(A)}_{\rm n}), 
\end{equation}
\begin{equation}\label{Eq:B_movement}
x^{(B)}\sim\mathcal{U}(-x^{(B)}_{\rm w},x^{(B)}_{\rm e}),\, y^{(B)}\sim\mathcal{U}(y^{(B)}_{\rm s},y^{(B)}_{\rm n}).
\end{equation}
\end{subequations}
In the latter expressions, $x^{(A)}_{\rm w}$, $x^{(A)}_{\rm e}$, $x^{(B)}_{\rm w}$, and $x^{(B)}_{\rm e}$ reveal the position limits in the $x$-axis for both nodes, while $y^{(A)}_{\rm s}$, $y^{(A)}_{\rm n}$, $y^{(B)}_{\rm s}$, and $y^{(B)}_{\rm n}$ represent their position limits in the $y$-axis. 

\section{Position Aided Beam Alignment }\label{sec:Position_BA}
Suppose that at the time instant $t+1$ both nodes $A$ and $B$ move to new positions represented by the points $M_{t+1}^{(A)}$ and $M_{t+1}^{(B)}$, respectively, with respective coordinates $(x_{t+1}^{(A)},y_{t+1}^{(A)})$ and $(x_{t+1}^{(B)},d_t-y_{t+1}^{(B)})$. According to the presented channel model, the channel between node $B$ and node $A$ at the discrete time instant $t+1$ can be expressed as $\mathbf{H}_{BA}(t+1) \triangleq \mathbf{A}_B\left(\boldsymbol{\theta}(t+1)\right){\rm diag}\{\boldsymbol{a}(t+1)\}\mathbf{A}_A^{\rm H}\left(\boldsymbol{\phi}(t+1)\right)$, with $\boldsymbol{\phi}(t+1)\triangleq[\phi_1(t+1)\,\phi_2(t+1)\,\cdots\,\phi_L(t+1)]$ and $\boldsymbol{\theta}(t+1)\triangleq[\theta_1(t+1)\,\theta_2(t+1)\,\cdots\,\theta_L(t+1)]$ denoting the AoDs and AoAs, respectively, resulting from the new node positions, while $\boldsymbol{a}(t+1)$ includes the channel gains. The beam alignment objective in this time instant is to design the BF vector at node $A$ and the combining vector at node $B$ as    
\begin{equation}\label{Eq:BA_Objective}
\left\{\mathbf{v}_A(t\!+\!1),\mathbf{u}_B(t\!+\!1)\right\}\!=\!\max_{\mathbf{f}\in\mathcal{F}_A,\mathbf{z}\in\mathcal{F}_B} \left|\mathbf{z}^{\rm H}\mathbf{H}_{BA}(t\!+\!1)\mathbf{f}\right|^2\!,
\end{equation} 
which implies BF gain maximization. If perfect knowledge of $\mathbf{H}_{BA}(t+1)$ was available at both nodes, nodes $A$ and $B$ should use the principal right and left singular vectors of $\mathbf{H}_{BA}(t+1)$ as the BF and combining vectors, respectively, to maximize the BF gain \cite{Love_TCOM_2003}. However, neither $\mathbf{H}_{BA}(t+1)$ for any $t$ can be estimated at any of the nodes due to the assumed hardware limitations nor the nodes are capable of realizing any arbitrary vector (they possess the finite vector sets $\mathcal{F}_A$ and $\mathcal{F}_B$). 

\subsection{Proposed Method}\label{sec:method}
Let us consider that, upon installation of the considered mmWave backhaul system at the initial discrete time instant $t$, both nodes $A$ and $B$ are aware of the coordinate system defined by the coordinates of the points $M_t^{(A)}$ and $M_t^{(B)}$. In the next discrete time instant $t+1$, suppose that both nodes move to the new position points $M_{t+1}^{(A)}$ and $M_{t+1}^{(B)}$. Each node's displacement in the $x$ and $y$ axes is assumed to be available to the node. The latter indicates that each node is aware of the coordinates of its new position, i$.$e$.$, node $A$ learns the coordinates $(x_{t+1}^{(A)},y_{t+1}^{(A)})$ and $B$ obtains $(x_{t+1}^{(B)},d_t-y_{t+1}^{(B)})$. If also the coordinates of the position of node $B$ become available to $A$ (through, for example, a dedicated control channel), then the latter node may estimate the AoD of the LOS channel path for its transmission at this time instant as 
\begin{equation}\label{Eq:New_phi}
\hat{\phi}_{1}(t+1) =
\left\{\!\!
\begin{array}{lr}
\arcsin\left(g_{t+1}\right),&x_{t+1}^{(B)}\geq x_{t+1}^{(A)}\\
\pi-\arcsin\left(g_{t+1}\right),\!&x_{t+1}^{(B)}<x_{t+1}^{(A)}\\
\end{array}
\right.,
\end{equation}
where the positive real $g_{t+1}$ is given by
\begin{equation}\label{Eq:g_value}
g_{t+1} = \frac{{\rm d}\left(O_{t+1,t+1},M_{t+1}^{(B)}\right)}{{\rm d}\left(M_{t+1}^{(A)},M_{t+1}^{(B)}\right)}
\end{equation}
with point $O_{t+1,t+1}$ having the coordinates $(x_{t+1}^{(B)},y_{t+1}^{(A)})$. From the specific nodes' positions at the time instant $t+1$ yields  
\begin{equation}\label{Eq:g_value}
g_{t+1} = \frac{d_{t}-y_{t+1}^{(A)}-y_{t+1}^{(B)}}{\sqrt{\left(x_{t+1}^{(A)}+x_{t+1}^{(B)}\right)^2 + \left(d_{t}-y_{t+1}^{(A)}-y_{t+1}^{(B)}\right)^2}}.
\end{equation}
In a similar way, if node $A$ shares its coordinates at the same time instant with $B$, then node $B$ can estimate the AoA of the LOS path for $A$'s transmission as 
\begin{equation}\label{Eq:New_theta}
\hat{\theta}_{1}(t+1) =
\left\{\!\!
\begin{array}{lr}
\pi + \phi_{1}(t+1),&x_{t+1}^{(B)}\geq x_{t+1}^{(A)}\\
2\pi - \phi_{1}(t+1),\!&x_{t+1}^{(B)}<x_{t+1}^{(A)}\\
\end{array}
\right..
\end{equation}
The AoD and AoA of the LOS path can be obtained similarly if node $B$ transmits and node $A$ operates in receive mode.
\begin{algorithm}[t!]\caption{Position Aided Beam Alignment}\label{Position_Algorithm}
\begin{algorithmic}[1]
\Statex \textbf{Initialization:} Construct a coordinate system upon installation of nodes $A$ and $B$ at the initial time instant $t$. Determine the minimum required SNR $\gamma_{o}$ for data communication.
\For{$n=t+1,t+2,\ldots$}
		\Statex{\textit{\textbf{Node $A$ Recovery Phase 1:}}}
		\If{$\gamma_n\geq\gamma_{o}$,}
			\State Set $\mathbf{v}_A(n)=\mathbf{v}_A(n-1)$, $\mathbf{u}_B(n)=\mathbf{u}_B(n-1)$, 
			\Statex \hspace{1.05cm}and transmit data.
		\Else
			\State Obtain the coordinates of the new position $M_{n}^{(A)}$.
			\State Compute the AOD $\hat{\phi}_{1}(n)$ using \eqref{Eq:New_phi} and the  
			\Statex \hspace{1.05cm}coordinates of the positions $M_{n}^{(A)}$ and $M_{n-1}^{(B)}$, and 
			\Statex \hspace{1.05cm}the point $O_{n,n-1}$. 
			\State Compute $\mathbf{a}_A(\hat{\phi}_1(n))$ and design $\mathbf{v}_A(n)$ using \eqref{Eq:v_vector}.
		\EndIf
		\If{$\gamma_n\geq\gamma_{o}$} 
			\State Set $\mathbf{v}_A(n)$ as in step $7$, $\mathbf{u}_B(n)=\mathbf{u}_B(n-1)$, 
			\Statex \hspace{1.05cm}and transmit data.
		\Else
			\State Send the coordinates of $M_{n}^{(A)}$ to Node $B$.
		\EndIf
		\Statex{\textit{\textbf{Node $B$ Recovery Phase:}}}
		\State Upon reception of a control signal with the coordinates  
		\Statex \hspace{0.55cm}of $M_{n}^{(A)}$, obtain the coordinates of $M_{n}^{(B)}$. 
		\State Compute the AOA $\hat{\theta}_{1}(n)$ using \eqref{Eq:New_theta} and the coordinates  
		\Statex \hspace{0.55cm}of the positions $M_{n}^{(A)}$ and $M_{n}^{(B)}$, and $O_{n,n}$. 
		\State Compute $\mathbf{a}_B(\hat{\theta}_1(n))$ and design $\mathbf{u}_B(n)$ using \eqref{Eq:u_vector}.
    \If{$\gamma_n\geq\gamma_{o}$,} 
			 \State Trigger node $A$ to transmit data.
			 \State Set $\mathbf{v}_A(n)$ as in step $7$, $\mathbf{u}_B(n)$ as in step $16$, 
			 \Statex \hspace{1.05cm}and transmit data.
    \Else
			\State Send the coordinates of $M_{n}^{(B)}$ to Node $A$ and use 
			\Statex \hspace{1.05cm}$\mathbf{u}_B(n)$ designed in step $16$.
		\EndIf
		\Statex{\textit{\textbf{Node $A$ Recovery Phase 2:}}}
		\State Upon reception of a control signal with the coordinates 
		\Statex \hspace{0.55cm}of $M_{n}^{(B)}$, compute the AOD $\hat{\phi}_{1}(n)$ using \eqref{Eq:New_phi} and the 
		\Statex \hspace{0.55cm}coordinates of $M_{n}^{(A)}$ and $M_{n}^{(B)}$, and $O_{n,n}$.
		\State Compute $\mathbf{a}_A(\hat{\phi}_1(n))$ and design $\mathbf{v}_A(n)$ using \eqref{Eq:v_vector}.
		\State Set $\mathbf{v}_A(n)$ as in step $23$, $\mathbf{u}_B(n)$ as in step $16$, 
		\Statex \hspace{0.55cm}and transmit data.
\EndFor
\end{algorithmic}
\end{algorithm}

Capitalizing on \eqref{Eq:New_phi} and on our channel model for each time instant $t+1$, we propose that transmit node $A$ uses its estimate $\hat{\phi}_{1}(t+1)$ to realize a beam steering towards the direction of receive node $B$. To accomplish this, it searches inside its beam codebook $\mathcal{F}_A$ for the vector that is closest to $\mathbf{a}_A(\hat{\phi}_1(t+1))$. In mathematical terms and without loss of generality, node $A$ designs its BF vector at each time instant $t+1$ as   
\begin{equation}\label{Eq:v_vector}
\mathbf{v}_A(t+1) = \min_{\mathbf{f}\in\mathcal{F}_A}\left\|\mathbf{f}-\mathbf{a}_A\left(\hat{\phi}_1(t+1)\right)\right\|_{\rm F}^2.
\end{equation}
Similarly, receive node $B$ uses its estimate $\hat{\theta}_{1}(t+1)$ at each $t+1$ to realize a beam as close as possible to the direction of node $A$. It, therefore, constructs its combining vector as  
\begin{equation}\label{Eq:u_vector}
\mathbf{u}_B(t+1) = \min_{\mathbf{z}\in\mathcal{F}_B}\left\|\mathbf{z}-\mathbf{a}_B\left(\hat{\theta}_1(t+1)\right)\right\|_{\rm F}^2.
\end{equation}
By using the vectors given by \eqref{Eq:v_vector} and \eqref{Eq:u_vector} at nodes $A$ and $B$, respectively, the instantaneous received SNR at node $B$ is given by $\gamma_{t+1} = \left|\mu_{t+1}\right|^2/\sigma_B^2$, where $\mu_{t+1}\in\mathbb{C}$ is defined as 
\begin{equation}\label{Eq:mu}
\begin{split}
\mu_{t+1} =& \sum_{\ell=1}^L\alpha_\ell(t+1)\mathbf{u}_B^{\rm H}(t+1)\mathbf{a}_B\left(\hat{\theta}_\ell(t+1)\right)   
\\&\times\mathbf{a}_A^{\rm H}\left(\hat{\phi}_\ell(t+1)\right)\mathbf{v}_A(t+1).
\end{split}
\end{equation}
It is noted that for the special case of $L=1$ LOS channel path, perfectly estimated AoD and AoD for this path, and infinite resolution beam codebooks at both nodes $A$ and $B$ yields $\mu_{t+1}=\alpha_1(t+1)$. This indicates that for this ideal case the vectors given by \eqref{Eq:v_vector} and \eqref{Eq:u_vector} maximize the BF gain described in \eqref{Eq:BA_Objective} and given by $\left|\mu_{t+1}\right|^2$. The proposed beam alignment method is summarized in Algorithm~\ref{Position_Algorithm}. 

\section{Numerical Results and Discussion}\label{sec:Results}
In this section we evaluate the performance of the proposed beam alignment method over the considered mmWave channel model. We particularly evaluate the OP performance defined as the probability that the instantaneous SNR falls below a minimum SNR threshold $\gamma_{o}$. To carry out this evaluation, we have simulated $10^4$ channel samples according to \eqref{Eq:Channel_Model} with normalized $\sigma_B^2$ and $P_L=d_1^{-3.75}$, where each channel sample appears at one discrete time instant. We have considered Ricean fading channels with the $\kappa$-factor denoting the ratio of the energy in the LOS channel path to the sum of the energies in the other non LOS paths \cite{Muhi-Eldeen}. The distance of the nodes $A$ and $B$ at the initial time instant $t=1$ is assumed to be $d_1=10$m and the random displacements of the nodes in the $x$-axis and $y$-axis due to wind gusts are obtained from \eqref{Eq:AB_movement} with $x^{(A)}_{\rm w}=x^{(A)}_{\rm e}=x^{(B)}_{\rm w}=x^{(B)}_{\rm e}=1.5$m and $y^{(A)}_{\rm w}=y^{(A)}_{\rm e}=y^{(B)}_{\rm w}=y^{(B)}_{\rm e}=1.5$m. The beam codebooks $\mathcal{F}_A$ and $\mathcal{F}_B$ are constructed by quantizing the feasible sets of departure and arrival angles, respectively. Specifically, for the $i$th BF vector belonging in $\mathcal{F}_A$ with $i\in{\rm card}(\mathcal{F}_A)$ the departure angle $\chi_i$ takes discrete values with step size $2^{1-q}(\chi_{\rm max}-\chi_{\rm min})$ within $[\chi_{\rm min},\chi_{\rm max}]$, where $q$ represents the number of angle quantization bits. Similarly, for each $j$th combining vector inside $\mathcal{F}_B$ with $j\in{\rm card}(\mathcal{F}_B)$ the arrival angle $\psi_j\in[\psi_{\rm min},\psi_{\rm max}]$ has been quantized with $q$ bits. We have set $\chi_{\rm min}=60^o$, $\chi_{\rm max}=120^o$, $\psi_{\rm min}=240^o$, and $\psi_{\rm max}=300^o$ with respect to the nodes' orientation. 
\begin{figure}[!t]
\centering
\includegraphics[width=0.432\textwidth]{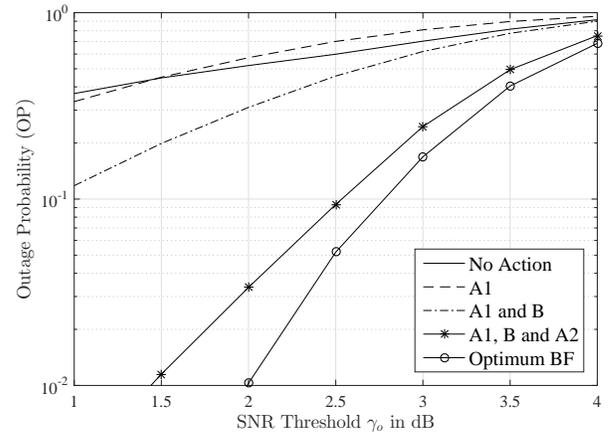}
\caption{OP versus the SNR threshold $\gamma_o$ in dB for $N_A=N_B=16$, $p/\sigma_B^2=5$dB, ${\rm card}(\mathcal{F}_A)={\rm card}(\mathcal{F}_B)=17$, $L=3$, $\kappa=13.2$dB, and different phases of the proposed method. Curves for the unfeasible optimum BF case and the case of no action for beam alignment are included.}
\label{Fig:Figure_3}
\end{figure}

In Fig$.$~\ref{Fig:Figure_3}, we plot the OP with the proposed beam alignment method as a function of the SNR threshold $\gamma_o$ in dB for the transmit SNR $p/\sigma_B^2=5$dB and $N_A=N_B=16$, ${\rm card}(\mathcal{F}_A)={\rm card}(\mathcal{F}_B)=17$ resulting from $q=5$, $L=3$, and the Ricean factor $\kappa=13.2$dB. Within this figures, the OP curves for the case of no action for beam alignment and the case of optimum BF \cite{Love_TCOM_2003} when perfect channel knowledge is available are also sketched. As for the proposed method, the performance for different sequences of phases is also demonstrated. In particular, we provide the performance for the case where only \textit{Node $A$ Recovery Phase 1} is used, denoted by ${\rm A}1$; the case where \textit{Node $A$ Recovery Phase 1} followed by \textit{Node $B$ Recovery Phase} are used, denoted by ${\rm A}1$ and ${\rm B}$; and the case where all phases are utilized, denoted by ${\rm A}1$, ${\rm B}$ and ${\rm A}2$. As seen and as expected, OP degrades with increasing $\gamma_o$. It is also shown that the exchange of position information improves this probability for any $\gamma_o$ value. In fact, the availability of nodes' position at both nodes, with only $2$ control signals according to the proposed method, results in the best performance. Evidently, actions from only the transmitter or exchange of only the latter's position information result in poor OP. We have also compared the proposed beam alignment method with an exhaustive beam search similar to \cite{Jeong_COMmag2015} that seeks to find the first beam pair meeting $\gamma_o$ whenever beam misalignment occurs. For the parameter settings of Fig$.$~\ref{Fig:Figure_3} we quote the following representative results: \textit{i}) the proposed method achieves $98\%$ of the average received SNR of optimum BF irrespective of $\gamma_o$, while the exhaustive search only the $28\%$ for $\gamma_o=1$dB after $30$ control signals on average; and \textit{ii}) for $\gamma_o=3$dB the exhaustive search reaches the $72\%$ of the optimum BF SNR requiring on average $153$ control signals.

\section{Conclusion}\label{sec:Conclusion}
In this paper, we investigated the problem of beam alignment between network nodes in wireless backhaul communication systems operating in the mmWave frequency band. We considered the practical case of random movement of the large phased antenna arrays of the nodes due to wind flow or ground vibration, and presented a simple model that captures their small displacements. A robust beam alignment method for wireless environments including a LOS component was presented that capitalizes on the exchange of position information between the nodes to design their BF and combining vectors. Through representative OP performance evaluation results the impact of some key parameters on the performance of the proposed method was highlighted. 

\bibliographystyle{IEEEtran}
\bibliography{IEEEabrv,refs}

\end{document}